\documentclass[english,twocolumn]{emulateapj}
\usepackage{helvet}
\usepackage[latin1]{inputenc}
\usepackage{graphicx}

\makeatletter

\providecommand{\tabularnewline}{\\}

\usepackage{natbib}
\usepackage{babel}
\makeatother
\begin{document}

\title{Source Matching in the SDSS and RASS: Which Galaxies are Really X-ray
Sources?}

\author{John K. Parejko, Anca Constantin, Michael S. Vogeley}

\affil{Department of Physics, Drexel University, Philadelphia, PA 19104}

\email{parejkoj@drexel.edu}

\and{}

\author{Fiona Hoyle}

\affil{Department of Physics and Astronomy, Widener University, Chester,
PA 19013}


\begin{abstract}
The current view of galaxy formation holds that all massive galaxies
harbor a massive black hole at their center, but that these black
holes are not always in an actively accreting phase. X-ray emission
is often used to identify accreting sources, but for galaxies that
are not harboring quasars (low-luminosity active galaxies), the X-ray
flux may be weak, or obscured by dust. To aid in the understanding
of weakly accreting black holes in the local universe, a large sample
of galaxies with X-ray detections is needed. We cross-match the ROSAT
All Sky Survey (RASS) with galaxies from the Sloan Digital Sky Survey
Data Release 4 (SDSS DR4) to create such a sample. Because of the
high SDSS source density and large RASS positional errors, the cross-matched
catalog is highly contaminated by random associations. We investigate
the overlap of these surveys and provide a statistical test of the
validity of RASS-SDSS galaxy cross-matches. SDSS quasars provide a
test of our cross-match validation scheme, as they have a very high
fraction of true RASS matches. We find that the number of true matches
between the SDSS main galaxy sample and the RASS is highly dependent
on the optical spectral classification of the galaxy; essentially
no star-forming galaxies are detected, while more than $0.6\%$ of
narrow-line Seyferts are detected in the RASS. Also, galaxies with
ambiguous optical classification have a surprisingly high RASS detection
fraction. This allows us to further constrain the SEDs of low-luminosity
active galaxies. Our technique is quite general, and can be applied
to any cross-matching between surveys with well-understood positional
errors.
\end{abstract}

\section{Introduction}

Distinguishing the processes that contribute to the emission from
the centers of galaxies is vital to understanding the co-evolution
of galaxies and their central black holes. Among nearby galaxies,
a large fraction of central emission sources are of ambiguous nature
\citep{1997ApJS..112..315H}; emission-lines in optical spectra of
many galaxies seem to reflect a mix of behavior between bona-fide
accretion (Seyfert-like) and active star formation (H II-like). In
order to discriminate between the various possible ionization mechanisms
and penetrate the obscuring dust layers that encircle these sources,
we need observations at multiple wavelengths. In particular, X-rays
are less prone to dust absorption and thus can be used to distinguish
between accretion sources and emission from young, hot stars. This
can clarify the observed optical emission spectra and allow us to
better describe the central accretion sources in low-luminosity active
galactic nuclei (AGN).

For an accurate census of the local galactic population, one must
study a statistically significant number of sources. The Sloan Digital
Sky Survey (SDSS, \citealp{2000AJ....120.1579Y}) provides the largest
sample of galaxies with spectra which allow emission-line classification
of central sources. The ROSAT All-Sky Survey (RASS, \citealp{1999A&A...349..389V,2000IAUC.7432....3V})
is the widest and deepest survey of the X-ray sky. The SDSS and RASS
are well matched in terms of depth, but have quite different astrometry
and spatial resolution. Previous studies matching a variety of SDSS
and RASS sources include analyzing the X-ray properties of spectroscopically
confirmed quasars \citep{2003AJ....126.2209A,2007AJ....133..313A},
generating an X-ray detected galaxy cluster catalog \citep{2004A&A...423..449P},
searching for optically unidentified neutron stars \citep{2006AJ....131.1740A},
and surveying the multi-wavelength properties of SDSS galaxies \citep{2006MNRAS.370.1677O}.

Because broad-line quasars are expected to be strong X-ray sources,
one would expect a large number of matches between RASS and SDSS for
these objects. \citet{2003AJ....126.2209A} characterized the RASS
properties of spectroscopically identified broad-line quasars from
the SDSS as well as some narrow-line sources. They qualitatively discuss
the likelihood that a given RASS-SDSS match is a true match and include
{}``normal'' (non or weakly emitting) galaxies as a comparison of
what a weak correlation would look like. They study more than $1000$
RASS-SDSS quasar/AGN and briefly discuss a few properties of the sample.
Their sample reproduces the expected non-linear optical/X-ray (2500\AA/2keV)
relationship among broad-line sources. The follow-up study, \citet{2007AJ....133..313A},
examines $\sim7000$ sources with similar results.

A different investigation involves identifying RASS sources with no
obvious optical counterpart. For example, this is useful for finding
optically dim neutron stars. \citet{2006AJ....131.1740A} identified
all SDSS sources within 4 times the positional error of each RASS
source. They then removed from their catalog any RASS source with
an SDSS match which could have produced the X-ray flux. After removing
objects with NED identifications, visually-identified bad fields,
and known galaxy clusters, 11 RASS sources with no plausible SDSS
optical counterpart remained. They claim this number is consistent
with the number of isolated neutron stars expected in the SDSS field.
Studying poorly understood matching samples in this way can clarify
whether the sample includes primarily true matches or primarily false
matches.

A recent comparison of RASS and SDSS in a multi-wavelength study \citep{2006MNRAS.370.1677O}
identified 267 RASS matches within $30\arcsec$ of SDSS DR1 main sample
galaxies \citep{2002AJ....124.1810S,2003AJ....126.2081A}. They list
a false association fraction of $\sim9\%$ (computed statistically
based on the RASS source density) and also show the positions of their
galaxies on an optical emission-line classification diagram \citep[the BPT diagram: ][]{1981PASP...93....5B}.
They did not investigate known-bad matches (as in \citealt{2006AJ....131.1740A}),
nor did they elaborate on the positions of the RASS detected galaxies
on their BPT diagram.

The ROSAT All Sky Survey was produced from data acquired in ROSAT's
scanning mode, but ROSAT also performed many individual targeted observations,
resulting in several pointed catalogs. These catalogs were generated
from serendipitous source discoveries made during individual targeted
observations. Because of this, they contain a large number of sources
in very small fields scattered over the sky with highly varying exposure
durations, making source upper limits difficult to compute. Previous
studies \citep[e.g.][]{2006AJ....132.1475S} have examined the properties
of SDSS quasars found in these catalogs.

\citet{2005mmgf.conf..320S} looked at star forming galaxies in the
SDSS DR1 and matched them to several different ROSAT catalogs, including
the RASS. Their final results involve 14 star forming galaxies which
they claim to be X-ray sources (four of which were previously studied).
We were not able to determine exactly which catalog they used in their
published results. Therefore, we cannot check whether the results
represent true matches between RASS and SDSS. Some star-forming galaxies
are expected to be X-ray emitters, but whether these galaxies are
actually detected in RASS remains to be seen.

The XMM-Newton and Chandra X-ray satellites both provide much improved
pointing, resolution and depth over ROSAT, but their fields of view
are quite small. Both have produced serendipitous source catalogs
similar to the ROSAT pointed catalogs mentioned above. The initial
XMM serendipitous source catalog was compared with the USNO A2.0 optical
catalog \citep{2006MNRAS.367.1017G} to find 46 optically identified
non-AGN galaxies with substantial X-ray flux. \citet{2005AJ....129...86H}
matched serendipitous source detections in Chandra with SDSS DR2 \citep{2004AJ....128..502A}
to find 42 X-ray emitting galaxies of a variety of types. The XMM-slew
survey \citep{freyberg-2005-} aims to solve the field of view and
uniformity problems by taking data during spacecraft slews between
targets. It will produce an all-sky map of equivalent depth to RASS,
with more than six times better resolution and pointing accuracy,
in roughly 6 years.

In this paper, we investigate the accuracy of matching RASS sources
with SDSS galaxies. In Section \ref{sec:Data} we describe the data
sets used in this study, including the systematics of selecting an
appropriate galaxy sample from SDSS. The details of the cross-matching
procedure and the statistical methods are described in Section \ref{sec:Cross-matching}
and the final matched data sets, separated by galaxy spectroscopic
class are detailed in Section \ref{sec:RASS-Detections-by}. We find
that a RASS/SDSS galaxy match cannot be trusted to represent the galaxy's
true X-ray flux without first identifying the galaxy's spectral type.
Section \ref{sec:Future-directions:-XMM-slew} provides a preliminary
analysis of the new XMM-slew catalog and shows its utility in clarifying
the presence of X-ray sources in galaxies.

\section{Data}

\label{sec:Data}

\subsection{SDSS}

This study employs data from the Sloan Digital Sky Survey Data Release
4 (SDSS DR4), an optical imaging and spectroscopic survey with spectroscopic
coverage of $\sim16\%$ of the sky as described in \citet{2000AJ....120.1579Y}
and \citet{2006ApJS..162...38A}. Technical details of the photometric
camera, telescope, analysis pipeline, monitor and related systems
can be found in \citet{1998AJ....116.3040G}, \citet{2006AJ....131.2332G},
\citet{2001ASPC..238..269L}, \citet{2001AJ....122.2129H} and \citet{2002AJ....123.2121S},
while \citet{2003AJ....125.1559P} describe the astrometric calibration
and \citealp{1996AJ....111.1748F} describe the u'g'r'i'z' photometric
system. The SDSS has very good photometric resolution ($\sim1\arcsec.4$
PSF) and astrometric precision ($<0\arcsec.1$ rms per coordinate)
\citep{2002AJ....123..485S}. The spectroscopic survey is constructed
from tilings of the photometric data \citep{2003AJ....125.2276B}
and includes the main galaxy sample, quasar sample and luminous red
galaxy sample which are described in \citet{2002AJ....124.1810S},
\citet{2002AJ....123.2945R}, and \citet{2001AJ....122.2267E} respectively.
This spectroscopic survey includes uniform, high quality spectra of
more than half a million galaxies and nearly 100,000 quasars, via
plates containing $3\arcsec$ diameter optical fibers.

\subsubsection{SDSS Galaxies}

Our focus is on the main galaxy spectroscopic sample which includes
all galaxies with Petrosian \emph{r} magnitudes brighter than 17.77
with the exception of those not observed due to fiber collision. Because
of the size of the fiber-plugs, spectroscopic targets for a single
plate must be separated by at least $55\arcsec$. This was more of
a problem in DR1 and DR2, before overlapping plates and follow-up
observations filled in many of the missing objects. The complete SDSS
spectroscopic catalog includes more galaxies with spectra than just
the main galaxy sample. We restrict ourselves to the main galaxy sample
to avoid sample bias; some SDSS objects were selected for spectroscopy
due to their proximity to FIRST radio sources \citep{1995ApJ...450..559B}
and/or RASS X-ray sources. See the appendix for details on our SDSS
source selection process, and the importance of using the main galaxy
sample in cross-matching studies.

SDSS studies at MPA/JHU produced a catalog%
\footnote{ Data catalogues from SDSS studies at MPA/JHU \url{http://www.mpa-garching.mpg.de/SDSS/}%
} of secondary source products generated from the SDSS spectroscopic
data \citep[see][for the DR2 catalog paper]{2004astro.ph..6220B}.
This catalog includes simultaneous measurements of the emission and
absorption line profiles. The complete catalog includes all objects
in SDSS (regardless of magnitude) that are spectroscopically identified
as galaxies; sources with emission-line widths greater than 1000km/s
are not included (thus all objects identified as Seyferts and LINERs
in this paper are type 2 objects). We restrict ourselves to the intersection
of the MPA/JHU catalog and the SDSS DR4 main galaxy sample described
above.

\subsubsection{\label{sub:Spectral-Classification-of}Spectral Classification of
Main Sample Galaxies}

\begin{figure}
\includegraphics[width=1\columnwidth]{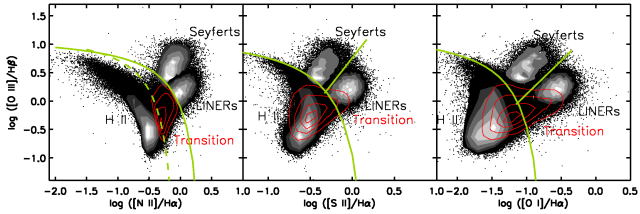}

\caption{\label{fig:emission-line classification diagram}Emission-line galaxy
classification diagram used to separate H IIs, Transitions, LINERs
and Seyferts. {}``Unclassified emission'' galaxies are those which
lie in a different region in each diagram.}
\end{figure}

We classify galaxies based on their optical emission-line properties.
Galaxies showing at least a $2\sigma$ detection of  flux in the emission-features
H$\alpha$, H$\beta$, {[}O III], {[}N II], {[}S II] and {[}O I] are
classified as emission-line galaxies, while those that show some but
not all of these lines are called {}``unclassifiable'' galaxies.
The strong line emitters are further separated into sources dominated
by accretion and those dominated by light from hot, young stars. We
classify H IIs, Seyferts, LINERs and Transition objects based on their
positions in a 4-dimensional space defined by the line-flux ratios
{[}O III]$\lambda5007$/H$\beta$, {[}N II]$\lambda6583$/H$\alpha$,
{[}S II]$\lambda\lambda6716,6731$/H$\alpha$, and {[}O I]$\lambda6300$/H$\alpha$.
We use the classification criteria from \citet{2006MNRAS.372..961K}.
Fig. \ref{fig:emission-line classification diagram} shows the regions
defining each galaxy subclass. We call those galaxies that do not
lie in the same classification region in each diagram, {}``unclassified
emission'' galaxies. Finally, galaxies showing no signs of emission
in H$\alpha$, H$\beta$ and {[}O III] are classified as {}``Passive''
galaxies. More details on this classification scheme can be found
in \citet{2007-Anca-inpress}.

\subsection{ROSAT All Sky Survey}

\begin{figure}
\includegraphics[angle=-90,width=1\columnwidth]{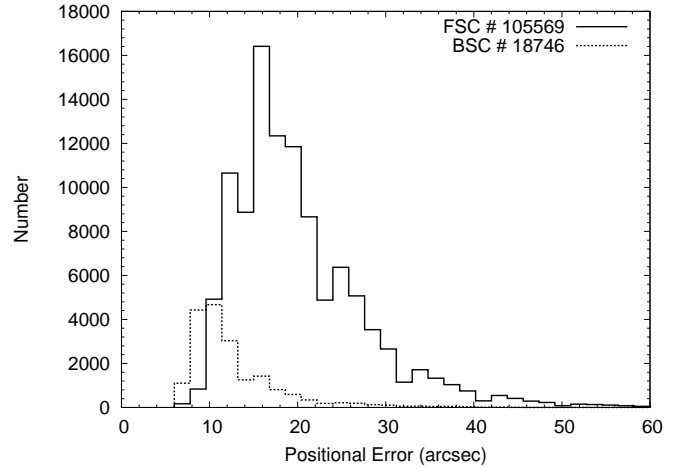}

\caption{\label{fig:ROSAT-positional-errors}RASS positional errors (rms)
in the Faint Source Catalog (FSC) and Bright Source Catalog (BSC).
The numbers after the hash (\#) in this, and all subsequent histograms,
give the total number of points included in that histogram.}
\end{figure}

Over the course of its eight year mission, the R\"{o}ntgensatellit
(ROSAT) produced a variety of distinct source catalogs from its two
X-ray detectors, the Position Sensitive Proportional Counter (PSPC)
and High Resolution Imager (HRI). The WGACAT \citep{2000yCat.9031....0W}
and 2RXP are serendipitous catalogs from pointed ROSAT observations
covering $\sim15\%$ and $\sim17\%$ of the sky, respectively. The
High Resolution Imager catalog (1RXH, \citealp{2000yCat.9028....0R})
covers $\sim2\%$ of the sky with much greater precision.

The PSPC scanning-mode data are the primary focus of this study: the
RASS Faint Source Catalog (FSC, \citealp{2000IAUC.7432....3V}) and
RASS Bright Source Catalog (BSC, \citealp{1999A&A...349..389V}) together
covering 92\% of the sky. We restrict ourselves to the RASS because
we would eventually like to compute source upper-limits. The average
integration time per target in the RASS varies between $<100$ seconds
for sources near the equator to $>5000$ seconds for sources near
the ecliptic poles, with $>97\%$ receiving more than 100 seconds.%
\footnote{RASS exposure map and ancillary data \url{http://www.xray.mpe.mpg.de/rosat/survey/rass-bsc/sup/}%
} 

The ROSAT PSPC operated between 0.2 and 2.4 keV, with the highest
sensitivity and resolution at roughly 1 keV. The PSPC optics were
focused for 1keV X-rays resulting in a $1\sigma$ PSF of roughly $25\arcsec$
at that energy. The resolution is worse for both higher energy (poor
focus) and lower energy (diffraction limit) X-rays. The scan-mode
observations that produced the RASS resulted in an astrometric positional
error ($1\sigma$ statistical error plus a $6\arcsec$ systematic
error) of 10-20'' (Fig. \ref{fig:ROSAT-positional-errors}). We show
in Section \ref{sub:Confirming-Quasars} that the $6\arcsec$ systematic
error is likely overestimated; $3\arcsec$ is likely more correct.

\subsection{XMM-Newton Slew Survey}

The RASS catalog is the current best compromise between width and
depth for X-ray data, but it has limitations, as noted above. To produce
an improved catalog, the X-ray Multi Mirror satellite (XMM-Newton)
is collecting X-ray counts during slews between targeted observations.
The first release of the XMM-Newton Slew Survey (XMM-slew, \citealp{freyberg-2005-})
covers 6240 square degrees of sky, in narrow north-south slews, using
the EPIC-pn CCD because of its large detector area, fast read-out
rate and high sensitivity to hard X-rays. Although average exposure
time is only $\sim10$s for any given source, the large mirror area
and sensitive detector make it nearly as deep as the RASS in the soft
band (0.2-2keV), and deeper and wider than any previous survey in
the hard band (2-12keV). The quoted $8\arcsec$ positional error along
the slew direction is dominated by the accuracy of the attitude reconstruction.
The EPIC-pn resolution of $4\arcsec$ is roughly a factor of 6 better
than the RASS resolution, thus XMM-slew can resolve many of the confused
RASS sources.

Two XMM-slew catalogs were released, a {}``total'' catalog containing
all detected sources, and a {}``clean'' catalog with known bad sources
removed and a higher detection threshold. We examine the clean sample
in this study; it contains 2713 sources with detections in at least
one band.

\section{Cross-matching}

\label{sec:Cross-matching}Cross-matching two surveys is simple enough:
count all objects separated by less than some threshold distance (in
our case, $60\arcsec$) as possible matches. But the validity of such
a match depends on the differing sky coverage, sensitivity, positional
accuracy and spatial resolution of the two matched surveys. These
differences lead to matches due to purely random associations, multiple
cross-matches for single sources, and erroneous flux measurements
due to contributions from multiple sources. For example, the ROSAT
PSPC is more than an order of magnitude worse than the SDSS in both
resolution and astrometry, and the SDSS source density is much higher.
Understanding the RASS-SDSS galaxy sample is particularly difficult
for sources that are not necessarily expected to be strong X-ray emitters,
such as spectroscopically identified low-luminosity narrow-line AGN,
passive or starburst galaxies. In this section, we attempt to quantify
the true and random components of RASS-SDSS cross-matches.

\begin{figure}
\includegraphics[width=1\columnwidth]{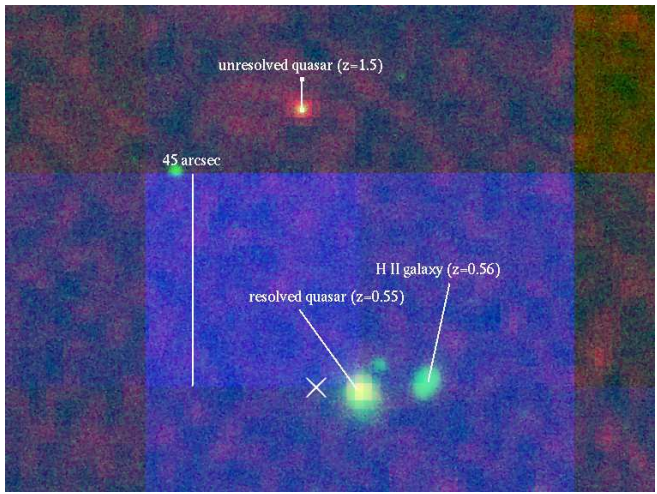}

\caption{\label{fig:2 Quasars and 1 H II}An example of RASS source confusion
(North is up). SDSS g-band is shown in green (spectroscopic classifications
are labeled), RASS pixels in blue (white X marks the source center)
and FIRST sources in red (both quasars are FIRST sources). Notice
that the RASS source covers two SDSS spectroscopic galaxies but is
centered on neither. There are many other cases where there is no
obvious source for the X-ray emission besides a single SDSS galaxy,
because of lack of SDSS spectroscopic information about all sources
in the field. The resolved quasar is SDSS J101643.87+421027.5 for
reference.}
\end{figure}

Fig. \ref{fig:2 Quasars and 1 H II} illustrates an example of the
issues faced in matching RASS and SDSS. Here a RASS source overlaps
two spectroscopically identified SDSS galaxies and is not centered
on either of them. One of the galaxies hosts a quasar and thus is
the likely source of the X-ray flux, while the other is identified
as a star-forming galaxy (H II-type optical spectrum) and thus is
expected to contribute little to the X-ray flux. If the quasar were
unidentified---because it had no spectrum taken---the star-forming
galaxy could have been considered the X-ray source. Another problem
is that the center of the X-ray source does not coincide with any
of the optical sources. This could be simply due to the astrometric
errors in the RASS catalog (Fig. \ref{fig:ROSAT-positional-errors}),
or to contributions to the total X-ray emission from the other quasar
at the top of the image. This example is not singular: there are many
such confusing matches in the RASS-SDSS galaxy sample because of the
high SDSS source density. Also, this RASS source is relatively bright,
and thus has better centroiding (positional error given as $8\arcsec$)
than most RASS sources and was particularly easy to catch.

\subsection{Obvious X-ray emitters}

\begin{figure}
\includegraphics[angle=-90,width=1\columnwidth]{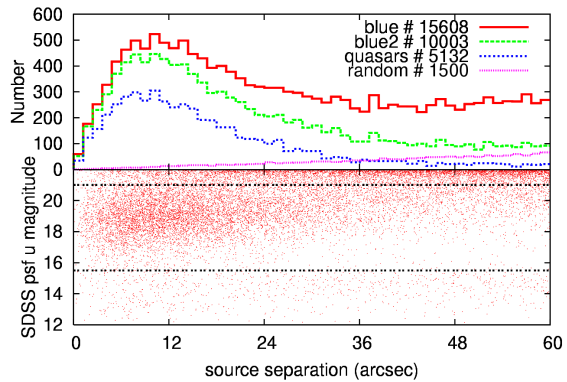}

\caption{\label{fig:SDSS-other-sources}Source-separation histogram for SDSS
sources that are expected to have a RASS identification. {}``Quasar''
includes spectroscopically identified quasars, {}``blue'' includes
point sources with $u-g<0.6$, {}``blue2'' is a subset of {}``blue''
restricted to $15.5<u<21$ and {}``random'' is a random match between
SDSS galaxies and RASS (see Section \ref{sub:Purely-random-matches}).
The lower plot shows the {}``blue'' sample, with the magnitude cuts
included. Note the vastly different tails for the three quasar-like
samples, implying a significantly different contamination fraction
in each.}
\end{figure}
When an SDSS object is the actual source of the RASS X-rays (a true
match), the distance between the X-ray and optical source positions
should be small. Some obvious choices for true matches are quasars
and quasar candidates. To qualitatively assess whether these {}``obvious''
choices are correct, we plot the distribution of distances between
the center of the RASS and SDSS sources---the source-separation histogram---for
these particular systems in Fig. \ref{fig:SDSS-other-sources}. The
upper panel includes the following RASS-matched SDSS sources: spectroscopically
identified quasars, sources with $u-g<0.6$ (quasar candidates), a
subset of the quasar candidates restricted to $15.5<u<21$ and {}``random
match'' between galaxies and the RASS, as described in the following
section. The lower panel shows the u-magnitude vs. source-separation
distribution for blue sources. Note the clustering of points at small
source-separations for $15.5<u<21$, suggesting that these are true
RASS-SDSS matches.

From the upper panel, spectroscopically identified quasars show an
obvious peak at small source-separations. {}``Blue'' objects (all
SDSS sources with $u-g<0.6$), which include some objects in the {}``quasar''
sample, have a peak at small separations as well as a prominent tail.
The {}``blue2'' sample (subset of {}``blue'' with $15.5<u<21$)
has a much smaller tail, suggesting a smaller fraction of incorrect
matches. Out of these samples, spectroscopically identified quasars
appear to represent the most reliable RASS-SDSS cross-match, with
the fewest points with large separations.

\subsection{Purely random matches}

\label{sub:Purely-random-matches}

Incorrect cross-matches between catalogs are due to random associations
between optical and X-ray sources. Previous work estimated the random
contamination by comparing the source density of the two catalogs,
which works well for samples with a small random contamination fraction.
We model these incorrect matches by generating {}``offset'' SDSS
object catalogs and matching them to the RASS. We produced 10 such
offset catalogs each from the SDSS galaxy and quasar catalogs by offsetting
all objects (either galaxies or quasars, respectively) from their
true RA and Dec by a fixed amount in a fixed direction, with a different
offset and direction for each offset catalog to reduce systematic
effects. The maximum offset was $1\arcdeg$ in RA and Dec. This procedure
preserves the on-sky source distribution of the SDSS, while moving
sources far away from their original RASS associations. When these
catalogs are matched to the RASS, the result is a linearly increasing
source-separation histogram, $\frac{dN}{dr}\propto r$; as the radius
increases, more sources fall within the matching circle. We compare
these random catalogs with our galaxy or quasar RASS matches to determine
the fractional contamination by purely random associations.

\subsection{\label{sub:Confirming-Quasars}Confirming Quasars}

\begin{figure}
\includegraphics[angle=-90,width=1\columnwidth]{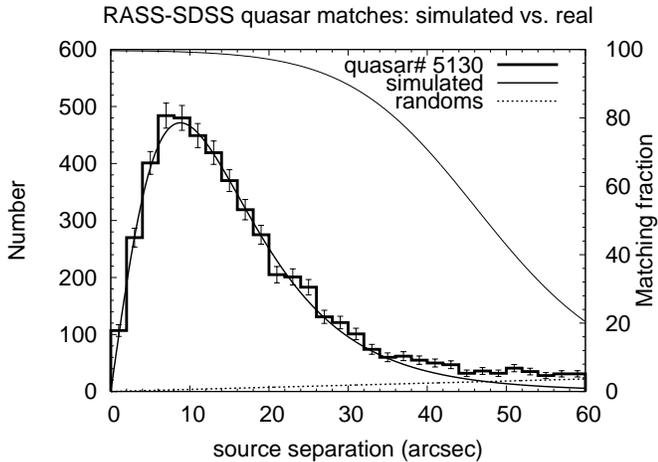}

\caption{\label{fig:Simulated-quasar-source}Simulated quasar source-separation
histogram for cross-matching between RASS and SDSS quasars. Note the
differing tails: the real distribution does not fall off as quickly,
as there is a small fraction of random matches at large radii. The
thin upper curve gives the true matching fraction at that radius (percent,
right axis).}
\end{figure}
X-ray source positional measurements have independent, normally distributed
errors in both planar components. This is analogous to darts thrown
at a small target. The precision of each throw is known, but individual
throws may have different precisions. The distribution of dart-target
distances is given by a Rayleigh distribution having a probability
density function (PDF), \[
P(r)\propto\frac{re^{-r^{2}/2\sigma^{2}}}{\sigma^{2}}\]
with scale parameter $\sigma$ and separation distance $r$. In the
case of X-ray measurements, the positional precision, $\sigma$ is
affected by the X-ray flux (reliability of centroiding depends on
the number of X-rays) and the pointing accuracy and resolution of
the measuring apparatus. The precision of each RASS source measurement
is listed in the catalog as the positional error (Fig. \ref{fig:ROSAT-positional-errors}).

We reproduce the source-separation histogram for RASS-SDSS quasar
matches by simulating X-ray source measurements using the corresponding
RASS positional errors plus a small random component. Because the
RASS positional errors are dependent on the X-ray flux, we use the
positional errors from the RASS-SDSS quasar matched catalog. For each
such RASS source, we generate a Rayleigh distribution with the positional
error of that source as the scale parameter $\sigma$. The sum of
the probability distribution function from each source gives our {}``simulated
true match'' curve. This PDF is the parent distribution for the true
matches between RASS and SDSS quasars. Random associations between
RASS and SDSS quasars have a linearly increasing source-separation
histogram, as shown above. A linear combination of these two distributions
(simulation PDF and random straight-line) should reproduce the observed
RASS-SDSS quasar source-separation histogram.

We show the quasar source-separation histogram, simulated true match
curve, and random component in Fig. \ref{fig:Simulated-quasar-source}.
The simulation curve, which does not include the random component,
matches the actual quasar source-separation histogram very well except
at the tail end. Combining the simulation and random components via
a $\chi^{2}$-minimization on the amplitude of each component yields
an excellent fit. The total fit is not shown in Fig. \ref{fig:Simulated-quasar-source}
because it would be completely masked by the data. This fit has a
$\chi^{2}$ per degree of freedom of $0.68$. However, the distributions
match only if the RASS positional errors are all reduced by $3\arcsec$,
implying that the quoted $6\arcsec$ systematic offset was overestimated.

The thin upper curve in Fig. \ref{fig:Simulated-quasar-source} gives
the {}``true matching fraction'' for RASS-SDSS quasar matches (percent,
right axis). This is the number of true matches (simulation curve)
divided by the total fit (simulation+random) at that radius. Note
that at $30\arcsec$, about 90\% of the RASS-SDSS quasars matches
are legimate. We also find that at $60\arcsec$ there is $\sim6\%$
total contamination to the RASS-SDSS quasar catalog. This agrees with
the estimate from \citet{2007AJ....133..313A} of $\sim5\%$ contamination
for their sample.

\subsection{Galaxies}

\begin{figure}
\includegraphics[angle=-90,width=1\columnwidth]{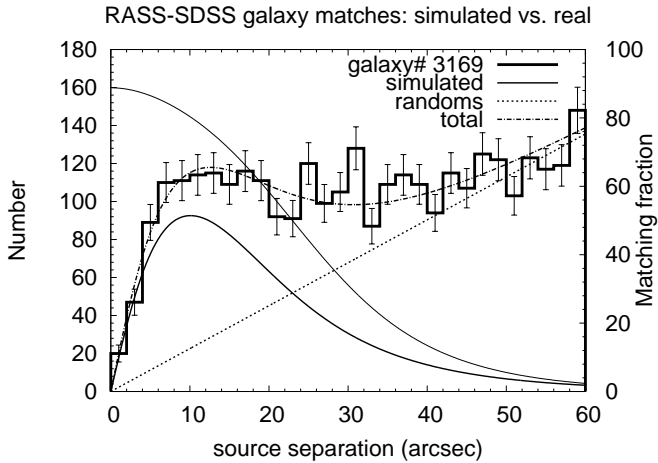}

\caption{\label{fig:Reconstructed-ROSAT/SDSS-galaxy}Reconstructed ROSAT/SDSS
galaxy source-separation with $\sim36\%$ true matches + $\sim64\%$
random matches out to $60\arcsec$. The thin curve from the upper
left to the lower right, cutting through the histogram, is the true
matching fraction (percent, right axis). There is very good agreement
between the total simulation curve and the actual distribution.}
\end{figure}
Matching SDSS main sample galaxies to the RASS results in 3169 total
matches. In contrast to quasars, the RASS-SDSS galaxy source-separation
histogram rises quickly, but is then relatively flat out to $60\arcsec$,
as seen in Fig. \ref{fig:Reconstructed-ROSAT/SDSS-galaxy}. This suggests
that while some galaxies are detected as X-ray sources, a large fraction
are simply random associations. We model the RASS-SDSS galaxy source-separation
histogram following the procedure outlined for quasars above. In this
case, the positional errors are those of the RASS-SDSS galaxy matched
catalog. Fig. \ref{fig:Reconstructed-ROSAT/SDSS-galaxy} compares
this model with the actual histogram. Note that the simulated true
match distribution is somewhat wider than the equivalent quasar curve,
as RASS sources associated with galaxies have a lower mean flux and
thus have larger positional errors. The $\chi^{2}$ per degree of
freedom of the total fit (simulated+random) is $1.18$ for galaxies.

To reduce the effect of source confusion in our RASS-SDSS galaxy sample,
we remove from our matched galaxy catalog RASS sources that are also
positionally matched with likely X-ray emitters. Our method is similar
to that employed by \citet{2006AJ....131.1740A} who removed RASS
sources that overlapped with spectroscopically identified quasars,
blue point sources (potential quasars), bright objects (ROSAT contaminant)
and sources with a quasar-like X-ray/optical spectral slope. Our requirements
are more relaxed, as our aim is not to eliminate all obvious x-ray
sources, but rather to identify X-ray counterparts of galaxies. Thus,
we only remove RASS sources from our matched galaxy catalog that are
close to the most reliable RASS cross-matches: within $40\arcsec$
of an SDSS quasar or within $30\arcsec$ of an object in the {}``blue2''
list described above. Also, if two SDSS galaxies match to one RASS
source, we take only the nearest match. This reduces the sample to
1970 galaxies, with many obviously incorrect matches removed, such
as the {}``match'' shown in Fig. \ref{fig:2 Quasars and 1 H II}.
This {}``cleaned'' catalog improves the $\chi^{2}$ of the simulation+random
fit to 0.96 and is the catalog employed in the analysis that follows.

\section{RASS Detections by Galaxy Spectroscopic Class}

\label{sec:RASS-Detections-by}

\begin{figure*}
\includegraphics[angle=-90,width=1\columnwidth]{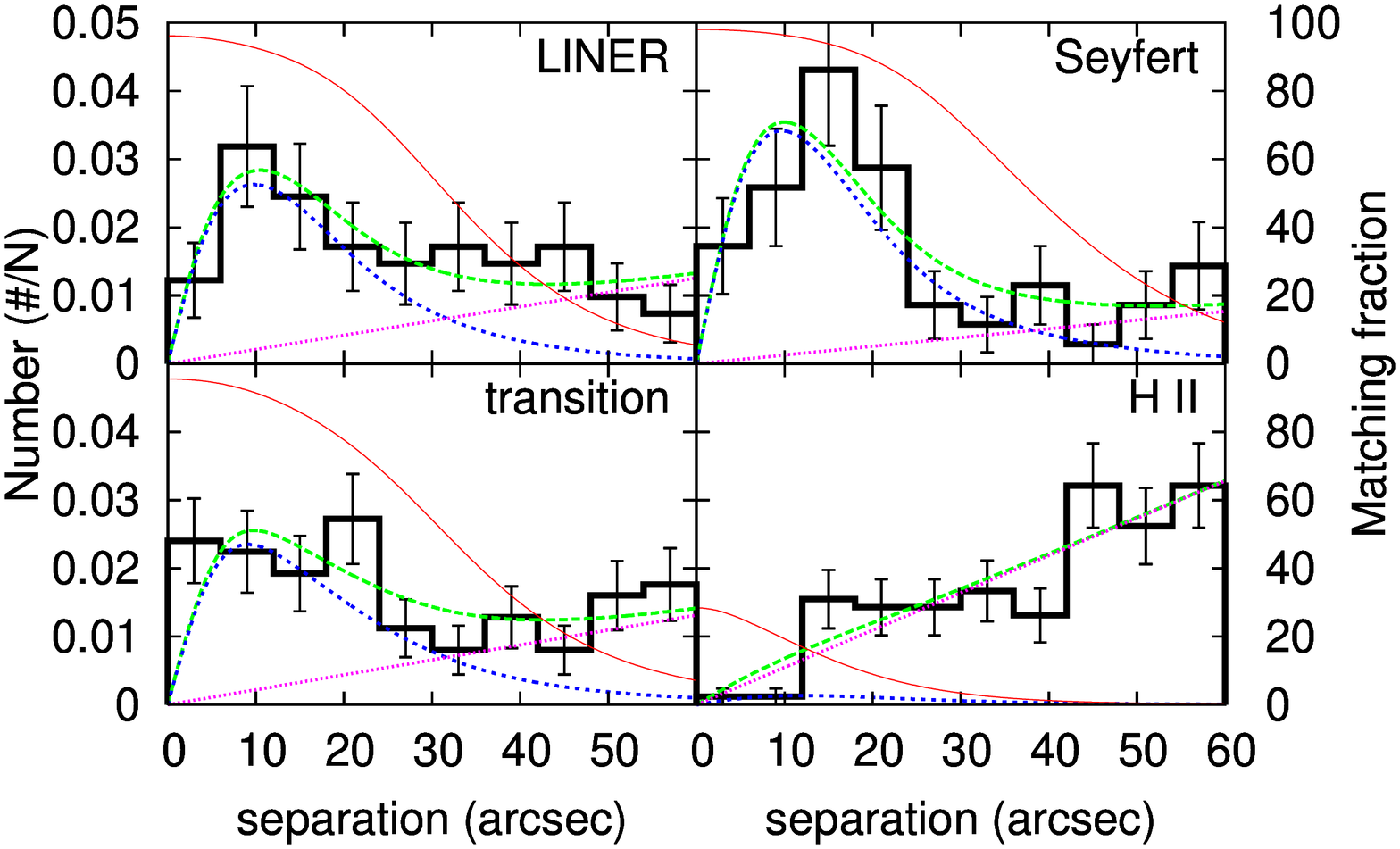}\hfil
\includegraphics[angle=-90,width=1\columnwidth]{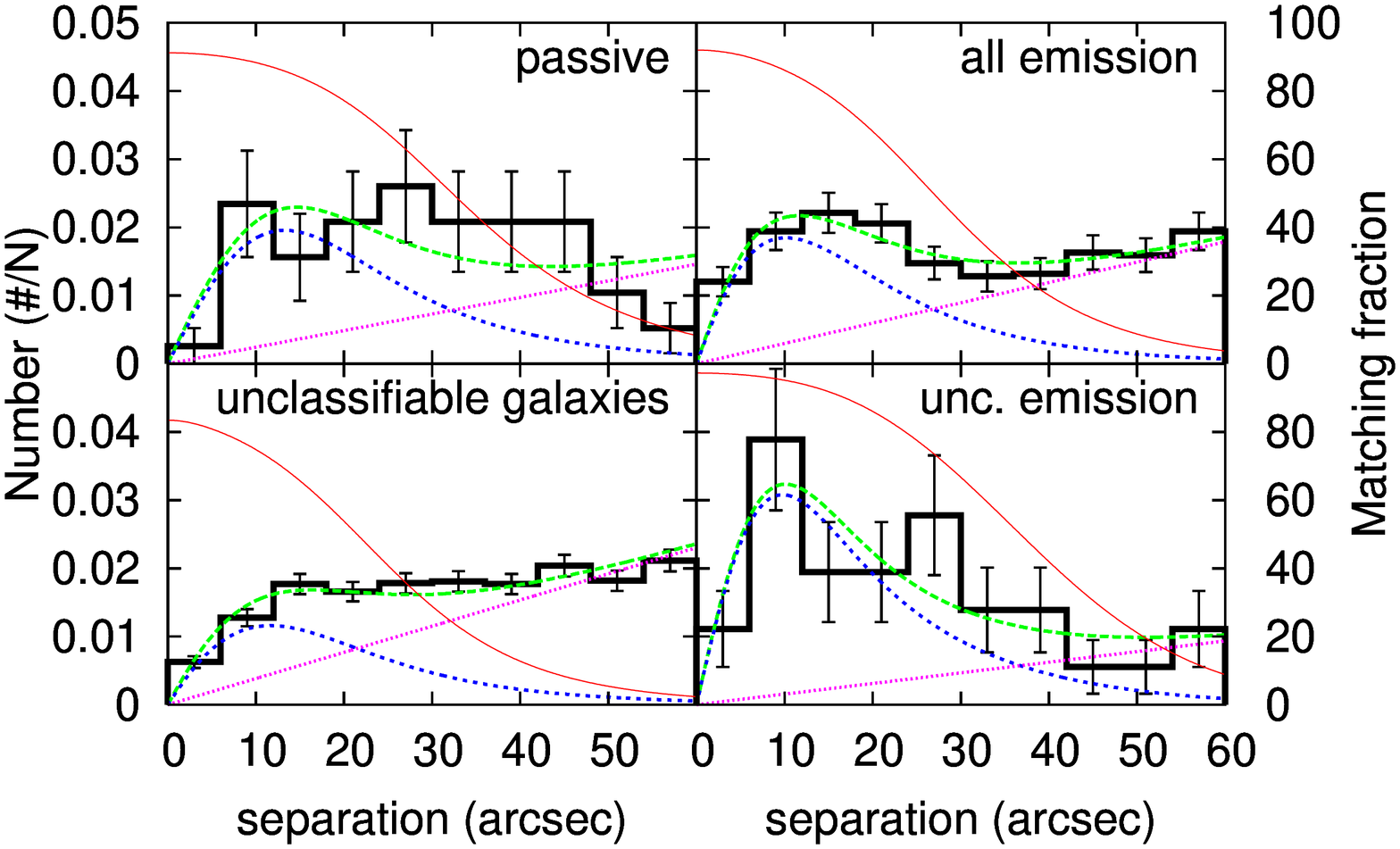}

\caption{\label{fig:ROSAT/SDSS_galaxy_subclass}ROSAT/SDSS galaxy source separation
by galaxy sub-type. Each sub-plot follows the structure of the individual
quasar and galaxy plots shown previously. The original matched distribution
(black, thick with $1\sigma$ Poisson errors), simulation (blue, short-dash)
and random (purple, dot) distributions produce the total simulated
distribution (green, long-dash). The thin solid curve (red, right
axis) gives the {}``true matching fraction'' described in Section
\ref{sub:Confirming-Quasars}.}
\end{figure*}

\begin{table*}
\tabletypesize{\scriptsize}
\begin{center}
\caption{\label{tab:chi-squared}$\chi^{2}$ per degree of freedom}

\begin{tabular}{cccccccccc}
\hline 
\hline&
all&
passive&
unclassifiable&
emission&
unc. emission&
H II&
transition&
LINER&
Seyfert\tabularnewline
\hline
$\chi^{2}$&
1.01&
1.24&
1.14&
0.47&
0.90&
1.53&
1.47&
0.47&
1.00\tabularnewline
\hline
\end{tabular}
\end{center}
\end{table*}
One would expect galaxies with different optical spectroscopic classes
to produce different X-ray fluxes and thus to have different matching
fractions. \citet{2006MNRAS.370.1677O} list the RASS matching fractions
for SDSS galaxies showing no emission as well as AGN, star-forming
and unknown emission-line galaxies. They also plot their RASS matches
on an emission-line classification diagram analogous to the left-most
plot in Fig. \ref{fig:emission-line classification diagram}. However,
they do not discuss random matches, nor do they remove known invalid
matches (e.g. quasars). Thus, their sample includes many SDSS galaxies
which are unlikely to be true matches to RASS sources. To investigate
the connection between RASS detection likelihood and optical spectroscopic
class, we separate the cleaned RASS-SDSS galaxy catalog into subclasses
as described in section \ref{sub:Spectral-Classification-of}. For
each of these subclasses, we simulate the source-separation histogram
via their corresponding RASS positional errors and linear random components
as before, and list the $\chi^{2}$ of the fits in Table \ref{tab:chi-squared}.

Fig. \ref{fig:ROSAT/SDSS_galaxy_subclass} compares the actual and
simulated distributions for the different galaxy classes. The left
plot shows the four different types of classified emission-line galaxies,
while the right plot shows the unclassified and passive galaxies.
The thin red curves show the true matching fraction at a given radius.
Note the high true matching fraction for galaxies with potentially
significant optical emission from a central accretion source: the
Seyfert, LINER and transition objects. Also note the relatively high
true matching fraction for unclassified emission and passive galaxies.
Galaxies with their optical emission dominated by star formation have
a very small true matching fraction; though there are a large number
of RASS-SDSS matches for H II and unclassifiable galaxies, most of
those matches are purely random associations.

\begin{table*}
\tabletypesize{\scriptsize}
\begin{center}
\caption{\label{tab:X-ray-detection-fractions}X-ray detection fractions (percent)}

\begin{tabular}{cccccccccc}
\hline 
\hline quasar&
all&
passive&
unclassifiable&
emission&
unc. emission&
H II&
transition&
LINER&
Seyfert\tabularnewline
\hline
8.3&
.12&
.41&
.11&
.11&
.28&
.004&
.19&
.41&
.66\tabularnewline
\hline
\end{tabular}
\end{center}
\end{table*}
We list the detection fractions for the various spectral classes in
Table \ref{tab:X-ray-detection-fractions}, including quasars for
comparison. This detection fraction is the integrated simulation curve
divided by the total number of galaxies in that class. Note the relatively
high detection fraction for galaxies with AGN-dominated optical emission,
including the tansition objects. The large X-ray detection fraction
for unclassified emission sources (defined in Section \ref{sub:Spectral-Classification-of})
suggests that many of these objects harbor obscured accretion.

The number of passive galaxy, unclassifiable galaxy, and LINER matches
to RASS are slightly under-predicted by the model at moderate radii
($20-40\arcsec$). Visual inspection of these galaxies confirms that
some of them are in or near clusters, which would produce an X-ray
source near to, but not coincident with, the galaxy. We do not have
a cluster catalog to remove these {}``contaminants'' but a visual
tally shows that between half and two-thirds of the RASS-matched passive
galaxies may be contaminated by the presence of a galaxy cluster.
However, some of these galaxies appear to be field galaxies, and thus
we may be finding X-ray bright, Optically Normal Galaxies \citep[XBONGs, see  ][]{2005MNRAS.358..131G}.
We plan to examine these objects in more detail in future work.

\begin{table*}
\tabletypesize{\scriptsize}
\begin{center}
\caption{\label{tab:Radius-within-which}Source-separation distance at fixed
true matching fraction}

\begin{tabular}{ccccccccccc}
\hline 
\hline fraction&
quasar&
all&
passive&
unclassifiable&
emission&
unc. emission&
H II&
transition &
LINER&
Seyfert\tabularnewline
\hline
85\%&
33&
6&
14&
\nodata&
11&
21&
\nodata&
16&
17&
23\tabularnewline
70\%&
40&
15&
24&
13&
25&
29&
\nodata&
24&
25&
30\tabularnewline
50\%&
47&
23&
32&
21&
32&
37&
\nodata&
32&
32&
38\tabularnewline
\hline
\end{tabular}
\tablecomments{ Listed here are the maximum
radii (expressed in arcseconds) for a given matching fraction, based
on the true match fraction shown in Fig. \ref{fig:ROSAT/SDSS_galaxy_subclass},
with quasars for comparison.}
\end{center}
\end{table*}
We list the RASS-SDSS source-separation radii at various fixed matching
fractions in Table \ref{tab:Radius-within-which} for all the objects
discussed in this paper. Notice that at no radius do H II galaxies
show even a $50\%$ true matching fraction. The true matching fraction
for star forming galaxies is extremely low because such galaxies do
not produce X-rays at a level detectable by the RASS and/or because
the X-rays they produce are completely obscured by dust. Because nearly
all RASS-SDSS star-forming galaxy matches are due to random associations,
no claims can be made about X-ray emitting star-forming galaxies from
these data alone. The XMM-Slew survey, XMM-Newton serendipitous source
catalog and the Swift BAT catalog all observe at higher X-ray energies
(less attenuated by dust), and so could help clarify the X-ray emission
properties of these galaxies.

\begin{table*}
\tabletypesize{\scriptsize}
\begin{center}
\caption{\label{tab:Percent-at-radius}Cumulative true matching fraction at
fixed radius}

\begin{tabular}{ccccccccccc}
\hline 
\hline radius&
quasar&
all&
passive&
unclassifiable&
emission&
unc. emission&
H II&
transition &
LINER&
Seyfert\tabularnewline
\hline
$40\arcsec$&
96.9&
53&
71&
48&
64&
84&
7&
75&
77&
88\tabularnewline
$30\arcsec$&
98.1&
64&
79&
59&
74&
89&
10&
83&
85&
92\tabularnewline
$20\arcsec$&
99.0&
75&
86&
70&
83&
94&
16&
89&
91&
95\tabularnewline
\hline
\end{tabular}

\tablecomments{ Listed here are the fractions
of each sample that are true matches, at the given radius, computed
by integrating the curves shown in Fig. \ref{fig:ROSAT/SDSS_galaxy_subclass},
with quasars for comparison.}
\end{center}
\end{table*}
For comparison with previous studies, we give the cumulative true
matching fraction at fixed radii in Table \ref{tab:Percent-at-radius}.
These values are computed from the ratio of the integrals of the simulated
and total curves in Fig. \ref{fig:ROSAT/SDSS_galaxy_subclass}, in
contrast with the previous table, derived from the point-wise ratios.
Again, note that H II galaxies have a very small cumulative true-match
fraction, even at small radii. All other matched sub-samples, except
for the unclassifiable galaxies, contain more than $85\%$ true RASS
matches below $20\arcsec$.

\section{Future directions: XMM-slew}

\label{sec:Future-directions:-XMM-slew}

\begin{figure}
\includegraphics[angle=-90,width=1\columnwidth]{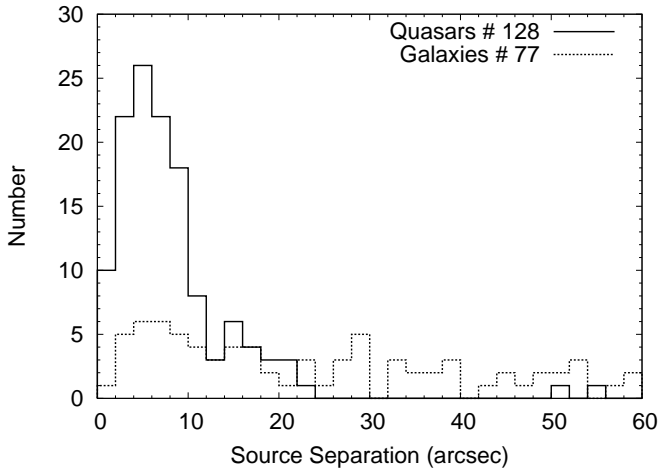}

\caption{\label{fig:XMM-slew}XMM-slew vs. SDSS cross matches for galaxies
and quasars. The $\sim8\arcsec$ XMM-slew positional errors are readily
visible in the quasar matches.}
\end{figure}
For comparison with RASS, we have matched the XMM-slew clean catalog
(first release) to both SDSS galaxies and quasars. Fig. \ref{fig:XMM-slew}
plots the source-separation histogram for these sources. The total
number of matches is quite small, due to the small number of XMM-slew
sources and the small overlap area between the surveys. Because of
the nature of the XMM-slew survey, we cannot perform the same analysis
as above; the narrow width of the slew strips is too small for a reliable
random fraction to be determined, yet. From the source-separation
histogram, $20\arcsec$ appears to be a reliable cut-off for true
matches. Accepting only those matches within this radius results in
$38$ galaxy matches and $115$ quasar matches to XMM-slew.

\begin{figure}
\includegraphics[width=1\columnwidth]{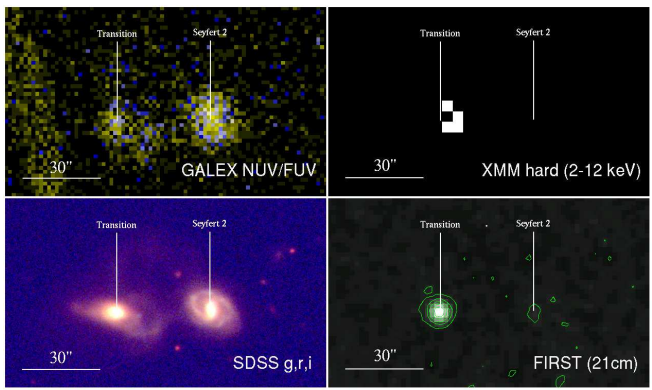}

\caption{\label{fig:SDSS-HII-in-XMM-slew}An interacting galaxy pair not detected
in soft X-rays. The central-source optical classifications are marked.
The SDSS optically-classified transition galaxy is found in the XMM-slew
survey at only hard (2-12keV) energies. The stripe on the left of
the GALEX image is due to detector edge effects. The pair is UGC 08327.}
\end{figure}
A coverage map for the XMM-slew data is not yet available, so it is
not possible to determine the percentage of ROSAT detections that
are non-detections in XMM-slew. However, amongst the $38$ {}``reliable''
matches are member(s) of each galaxy class described above. Most of
these XMM-slew detections are in the soft band ($0.2-2$ keV), but
there are a few galaxies with a detected hard X-ray flux. An example
is shown in Fig. \ref{fig:SDSS-HII-in-XMM-slew}: an interacting pair
of galaxies optically classified as a Transition and a Seyfert. The
Transition galaxy shows a hard X-ray flux and substantial radio point
source, while the Seyfert is unidentified in hard and soft X-rays
and shows a $\sim2\sigma$ detection in the FIRST catalog. There is
no RASS source at this location. We plan to followup on this intriguing
pair to better understand their emission properties and spectral shape.

\section{Conclusions}

We have examined the matching statistics between the ROSAT All Sky
Survey and the SDSS main galaxy sample. Our technique---simulating
the RASS-SDSS source-separation via the RASS positional errors plus
a linear random component---can reproduce the measured source-separations
for RASS-SDSS quasar matches as well as RASS-SDSS galaxies and subclassifications
of galaxies. We find that the likelihood of a given cross-match match
being a true match depends strongly on the optical spectral classification
of a given galaxy. We find that essentially no optically classified
star-forming galaxy has a true RASS counterpart, while LINERs, Seyfert
2s and Transition and unclassified emission galaxies do have reliable
X-ray detections. We also find a surprising number of galaxies lacking
optical emission lines which appear to be detected in the RASS. A
complete, low-redshift SDSS galaxy cluster catalog could be used to
clarify these XBONG candidates.

Our technique can be applied to any cross-matching between two surveys.
The only requirement is that the positional errors of each measurement
be known; no arbitrary fitting parameters are needed. By comparing
the observed source-separation histogram with a linear combination
of the probability distribution functions computed from the positional
errors and a random matched catalog, a {}``true matching fraction''
can be determined for any two matched catalogs. This is not limited
to X-rays: as a test, we were also able to reproduce the source-separation
histogram for a matched catalog of SDSS spectroscopic stars and GALEX
UV sources. The technique works best for catalogs containing mostly
point sources, as centroiding extended sources can be difficult and
the centers of sources may be wavelength-dependent.

\acknowledgements{We are grateful to Jenny Greene, Alina Badus and Danny Pan for providing
useful comments on earlier drafts of this paper. We acknowledge support
from NASA grant NAG5-12243 and NSF grant AST-0507647.\\
Funding for the Sloan Digital Sky Survey (SDSS) and SDSS-II has been
provided by the Alfred P. Sloan Foundation, the Participating Institutions,
the National Science Foundation, the U.S. Department of Energy, the
National Aeronautics and Space Administration, the Japanese Monbukagakusho,
and the Max Planck Society, and the Higher Education Funding Council
for England. The SDSS Web site is http://www.sdss.org/.\\
The SDSS is managed by the Astrophysical Research Consortium (ARC)
for the Participating Institutions. The Participating Institutions
are the American Museum of Natural History, Astrophysical Institute
Potsdam, University of Basel, University of Cambridge, Case Western
Reserve University, The University of Chicago, Drexel University,
Fermilab, the Institute for Advanced Study, the Japan Participation
Group, The Johns Hopkins University, the Joint Institute for Nuclear
Astrophysics, the Kavli Institute for Particle Astrophysics and Cosmology,
the Korean Scientist Group, the Chinese Academy of Sciences (LAMOST),
Los Alamos National Laboratory, the Max-Planck-Institute for Astronomy
(MPIA), the Max-Planck-Institute for Astrophysics (MPA), New Mexico
State University, Ohio State University, University of Pittsburgh,
University of Portsmouth, Princeton University, the United States
Naval Observatory, and the University of Washington\\
We have made use of the ROSAT Data Archive of the Max-Planck-Institut
für extraterrestrische Physik (MPE) at Garching, Germany.}

\appendix{\label{app:SDSS-galaxy-selection}}

\section{SDSS galaxy selection}

The main galaxy sample does not contain all the galaxies with spectra:
a galaxy could also have a spectrum taken if it is within $2\arcsec$
of a FIRST radio source or within the error-circle ($10-30\arcsec$)
of a RASS source. Luminous red galaxies are selected for follow-up
spectra based on their position in the (g-r, r-i, i) color-color-magnitude
cube. Spectra are also taken for a variety of serendipitous sources
including low surface-brightness galaxies. These other sources are
all dimmer than 17.77 in the r-band, and biased toward AGN and star-forming
galaxies. The systematics of these serendipitous sources are poorly
understood.

\begin{figure}
\includegraphics[angle=-90,width=1\columnwidth]{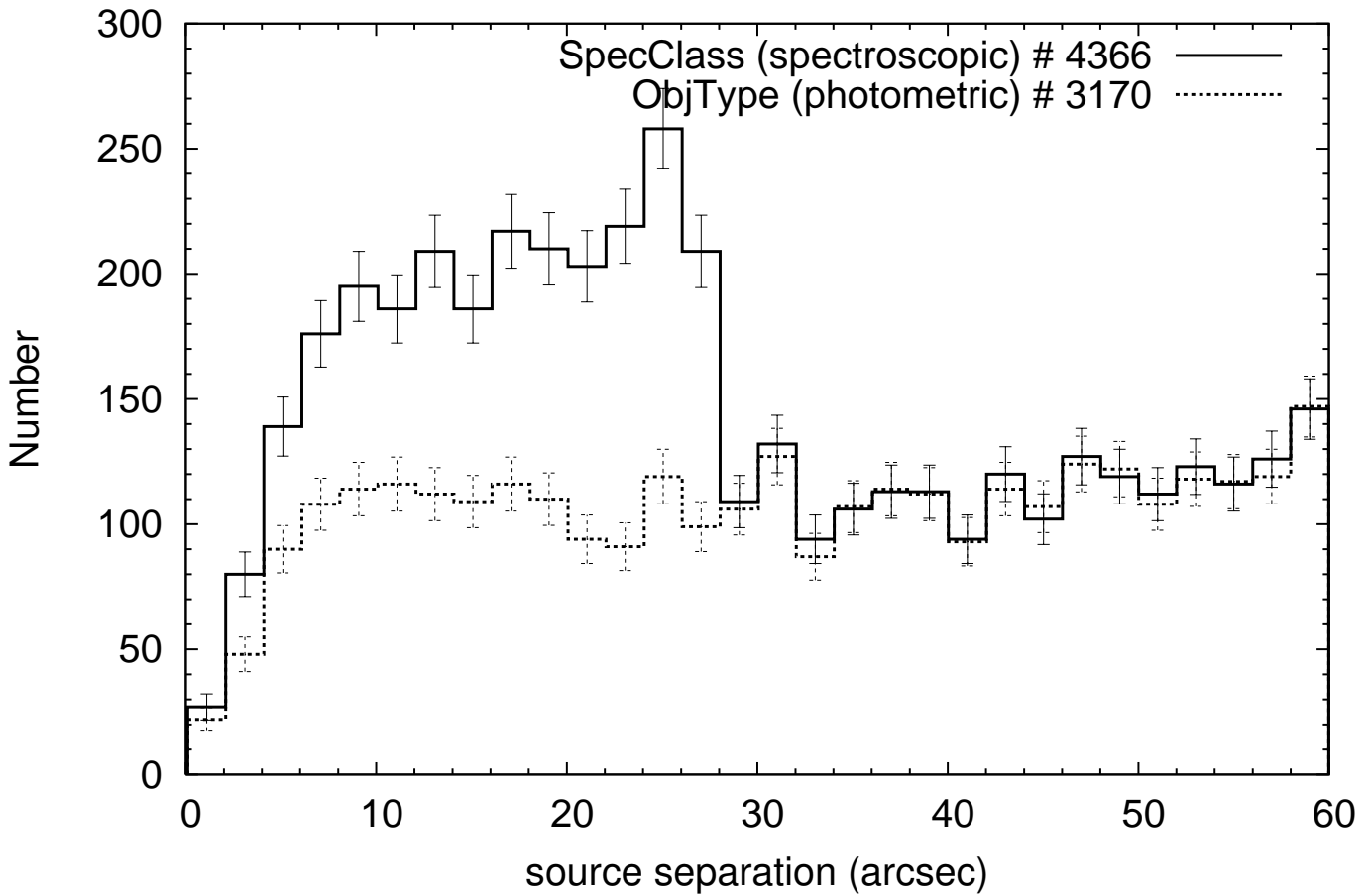}

\caption{\label{fig:RASS_photo_vs_spectro}SDSS Galaxies with RASS matches
within 1'. Note the sharp drop at $30\arcsec$ within the sample of
objects spectroscopically classified as galaxies compared to the main
galaxy sample (all galaxies with $r_{petro}<17.77$). Some galaxies
which are not in the main sample were targeted specifically because
they were within $30\arcsec$ of a RASS source.}
\end{figure}
The primary method for downloading large data sets from SDSS is CasJobs%
\footnote{SDSS CasJobs website \url{http://casjobs.sdss.org/CasJobs/}%
}. To extract the main galaxy sample from SDSS CasJobs, use the SpecObj
parameter ObjType and select those objects classified as {}``Galaxy''.
This includes all objects that were targeted for spectroscopy because
they met the main galaxy sample criterion. This classification is
\emph{before} the spectra were taken, and is thus a uniform sample.
A more na\"{\i}ve selection might be to take all objects spectroscopically
classified as galaxies: those with SpecObj parameter SpecClass listed
as {}``Galaxy''. However, this sample includes all objects with
a galaxy-like spectrum, which includes objects targeted for the above
reasons in addition to the main galaxy sample.

In Fig. \ref{fig:RASS_photo_vs_spectro} we show the source-separation
histogram for these two different samples. The {}``photometric''
sample is the main galaxy sample used in this study. The spectroscopic
sample, with a peak at $30\arcsec$, includes objects specifically
targeted because they were near a RASS source. The fiber-selection
process allocates spare spectroscopic fibers to sources within $30\arcsec$$ $
of a RASS source. These objects, having SpecObj parameter ObjType
classifications {}``ROSAT\_A'', {}``ROSAT\_B'', {}``ROSAT\_C''
or {}``ROSAT\_D'' account for roughly $2\%$ of all objects with
spectra in SDSS. \citet{2002AJ....123..485S} claim over half of these
ROSAT-based targets turn out to be quasars or AGN. This results in
a factor of two increase in potential matches at matching radii below
30''. This is why a statistical analysis of RASS matches to SDSS
must stick with the main galaxy sample; the other sources were selected
non-uniformly, and though they may result in odd and interesting spectra,
they produce a strong bias in X-ray matching properties.

\bibliographystyle{apj}
\bibliography{bib/xray,bib/sdss,bib/sdss-tech,bib/radio,bib/voids,bib/agn-general}

\end{document}